\documentclass[final,1p,times,twocolumn]{elsarticle}

\usepackage{amssymb}

\journal{Physics of the Dark Universe}

\begin{document}

\begin{frontmatter}



\title{Gravitational wave astronomy and cosmology}


\author{Scott A.\ Hughes}

\address{Department of Physics and MIT Kavli Institute, 77 Massachusetts
Avenue, Cambridge, MA 02139}

\begin{abstract}
The first direct observation of gravitational waves' action upon
matter has recently been reported by the BICEP2 experiment.  Advanced
ground-based gravitational-wave detectors are being installed.  They
will soon be commissioned, and then begin searches for high-frequency
gravitational waves at a sensitivity level that is widely expected to
reach events involving compact objects like stellar mass black holes
and neutron stars.  Pulsar timing arrays continue to improve the
bounds on gravitational waves at nanohertz frequencies, and may detect
a signal on roughly the same timescale as ground-based detectors.  The
science case for space-based interferometers targeting millihertz
sources is very strong.  The decade of gravitational-wave discovery is
poised to begin.  In this writeup of a talk given at the 2013 TAUP
conference, we will briefly review the physics of gravitational waves
and gravitational-wave detectors, and then discuss the promise of
these measurements for making cosmological measurements in the near
future.
\end{abstract}


\end{frontmatter}


\section{Introduction and overview}
\label{sec:intro}

Although often introduced as a consequence of Einstein's theory of
general relativity (GR), gravitational radiation is in fact necessary
in any relativistic theory of gravity.  These waves are simply the
mechanism by which changes in gravity are causally communicated from a
dynamical source to distant observers.  In GR, the curvature of
spacetime (which produces tidal gravitational forces) is the
fundamental field characterizing gravity.  Gravitational waves (GWs)
are propagating waves of spacetime curvature, tidally stretching and
squeezing as they radiate from their source into the universe.

Tidal fields are quadrupolar, so GWs typically arise from some
source's bulk, quadrupolar dynamics.  Consider a source whose mass and
energy density are described by $\rho$.  Choosing the origin of our
coordinates at the source's center of mass, its quadrupole moment is
given by
\begin{equation}
Q_{ij} = \int\rho\left(x_i x_j - \frac{1}{3}\delta_{ij}r^2\right)\,dV\;,
\label{eq:Qij}
\end{equation}
where the integral is taken over the source.  The gravitational-wave
potential, $h_{ij}$, comes from the second time derivative of
$Q_{ij}$:
\begin{equation}
h_{ij} = \frac{2G}{c^4}\frac{1}{r}\frac{d^2Q_{ij}}{dt^2}\;,
\label{eq:hij}
\end{equation}
where $r$ is distance from the source to the observer.  The magnitude
of a typical component of $h_{ij}$ is
\begin{equation}
h \approx \frac{G}{c^4}\frac{mv^2}{r}\;,
\end{equation}
where $v$ is the typical speed associated with the source's
quadrupolar dynamics, and $m$ is the mass that participates in those
dynamics.  Notice the combination of constants appearing here,
\begin{equation}
\frac{G}{c^4} = 8.27\times10^{-50}\,{\rm gm}^{-1}{\rm
    cm}\left(\frac{\rm cm}{\rm sec}\right)^{-2}\;.
\end{equation}
This is rather small, reflecting the fact that gravity is the weakest
of the fundamental forces.  To overcome it, one must typically have
large masses moving very quickly.  A short-period binary in which each
member is a compact object (white dwarf, neutron star, or black hole)
is a perfect example of a strong quadrupolar radiator.  For many of
the sources we discuss, $m$ is of order solar masses (or even millions
of solar masses), and $v$ is a substantial fraction of the speed of
light.

(In addition to quadrupole dynamics, there is one other well-known
mechanism for producing GWs: the amplification of primordial
ground-state fluctuations by rapid cosmic expansion.  We will briefly
discuss this way of producing GWs in Sec.\ {\ref{sec:ultra}}.)

The GWs a source emits backreact upon it, which appears as a loss of
energy and angular momentum.  The ``quadrupole formula'' predicts that
a system with a time changing quadrupole moment will lose energy to
GWs according to
\begin{equation}
\frac{dE}{dt} = -\frac{1}{5}\frac{G}{c^5}\sum_{ij}\frac{d^3Q_{ij}}{dt^3}
\frac{d^3Q_{ij}}{dt^3}\;.
\end{equation}
This loss of energy from, for example, a binary star system will
appear as a secular decrease in the binary's orbital period ---
orbital energy is lost to GWs and the stars fall closer together.
This effect was first seen in the first known binary neutron star
system, PSR 1913+16 (the famed ``Hulse-Taylor'' pulsar) {\cite{ht}}.
In this system, each neutron star has a mass slightly over
$1.4\,M_\odot$, and they orbit each other in less than 8 hours.  The
period has been observed to decrease by about 40 seconds over a
baseline of nearly 40 years of observation.  Similar period evolutions
have now been measured in about 10 galactic binaries containing
pulsars {\cite{inspiraling}}, and has even been seen in optical
measurements of a close white dwarf binary system
{\cite{wd_inspiral}}.

From these ``indirect'' detections, a major goal now is to directly
detect GWs.  Except at the longest wavelengths (where direct detection
has recently been reported), almost all measurement schemes use the
fact that a GW causes oscillations in the time of flight of a light
signal; the basic idea was sketched by Bondi in 1957 {\cite{bondi}}.
Imagine an emitter located at $x = x_e$ that generates a series of
very regular pulses, and a sensor at $x = x_s$.  Ignoring the nearly
static contribution of local gravitational fields (e.g., from the
Earth and our solar system), the spacetime metric through which the
light pulses travel can be written
\begin{equation}
ds^2 = -c^2dt^2 + \left[1 + h(t)\right] dx^2\;.
\end{equation}
Light moves along a null trajectory for which $ds^2 = 0$, which means
that that the speed of light with respect to these coordinates is
(bearing in mind that $h \ll 1$)
\begin{equation}
\frac{dx}{dt} = c\left[1 - \frac{1}{2}h(t)\right]\;.
\end{equation}
The time it takes light to travel from the emitter to the receiver is
\begin{equation}
\Delta T = \int_{x_e}^{x_s}\frac{dx}{dx/dt}
= \frac{x_s - x_e}{c} + \frac{1}{2c}\int_{x_e}^{x_s}h(t)\,dx\;.
\end{equation}
The gravitational wave thus enters as an oscillation in the arrival
time of pulses.  If the emitter is regular enough to be a precise
clock, one may measure the GW by measuring this oscillation.

Before rushing out to build our detector, we should estimate how
strong the gravitational waves we seek are.  We use the formula for
$h$ given above, substituting fiducial values for the physical
parameters that are likely to characterize the sources we aim to
measure:
\begin{eqnarray}
h &\approx& \frac{G}{c^4}\frac{mv^2}{r}
\nonumber\\
&\approx& 10^{-22}\times\left(\frac{200\,{\rm Mpc}}{r}\right)\times
\left(\frac{M}{3M_\odot}\right)\times \left(\frac{v}{0.3c}\right)^2
\nonumber\\
&\approx& 10^{-20}\times\left(\frac{6\,{\rm Gpc}}{r}\right)\times
\left(\frac{M}{10^6\,M_\odot}\right)\times \left(\frac{v}{0.1c}\right)^2\;.
\end{eqnarray}
The first set of numbers characterizes stellar mass sources that are
targets for ground-based high-frequency detectors, discussed in
Sec.\ {\ref{sec:high}}; the second characterizes massive black holes
that are targets of space-based low-frequency detectors discussed in
Sec.\ {\ref{sec:low}}, and (at somewhat higher $M$, lower $v$, and
smaller $r$) of pulsar timing arrays discussed in
Sec.\ {\ref{sec:very}}.

The numbers for $h$ are {\it tiny}.  Measuring timing oscillations at
this level of precision might seem crazy.  However, there is no issue
of principle that prevents us from measuring effects at this level;
the real challenge is to ensure that noise does not obscure the signal
we hope to measure.  Recall that a gravitational wave acts as a tidal
force.  The tide per unit mass for a GW of amplitude $h$ and frequency
$\omega$ is $R \simeq \omega^2 h$.  Considering a light source and
sensor separted by distance $L$, this means that we must control
against stray forces on our test mass $m$ of magnitude
\begin{equation}
F \simeq mL\omega^2 h \simeq 6\,{\rm
  piconewtons}\left(\frac{m}{40\,{\rm kg}}\right)
\left(\frac{f}{100\,{\rm Hz}}\right)^2 \left(\frac{L}{4000\,{\rm
    m}}\right) \left(\frac{h}{10^{-22}}\right)\;.
\end{equation}
(These fiducial parameters correspond to the LIGO observatories.)  Six
piconewtons is small, but it is well within our reach to isolate
against forces of this magnitude --- this is roughly the weight of a
single animal cell.  Though challenging, measuring a GW of $h \sim
10^{-22}$ is within our grasp.

In the remainder of this article, we discuss some of the science of
GWs.  We break up our discussion by frequency band.  We begin with the
{\it high frequency} band, with wave frequencies ranging from Hz to
kHz, which are targeted by ground-based interferometers; then move to
{\it low frequency}, waves with periods of minutes to hours, which are
targets of space-based interferometers; then {\it very low frequency},
waves with periods of order months to years, which are targets of
pulsar timing arrays; and finally conclude with {\it ultra low
  frequency}, with wavelengths comparable to the size of the universe.

\section{The spectrum of gravitational waves}

\subsection{High frequency}
\label{sec:high}

The high-frequency band of roughly $1 - 1000$ Hz is targeted by
ground-based laser interferometers.  The lower end of this band is set
by gravitational coupling to local seismic disturbances, which can
never be isolated against {\cite{hughesthorne}}; the upper end is set
by the fact that 1 kHz is roughly the highest frequency that one
expects from astrophysical strong GW sources.  In laser
interferometry, the laser's very stable frequency serves as the clock
for the measurement procedure sketched in Sec.\ {\ref{sec:intro}}.
GWs are detected by their action on light progating between widely
separated (hundreds to thousands of meters) test masses.

Several facilities around the world are involved in the search for
GWs.  Some of these facilities are presently offline as they undergo
upgrades to ``advanced'' sensitivity, but will begin active GW
searches again in about two years.  There is very close collaboration
among the facilities' research groups; combining data from multiple
observatories greatly increases the ability to discriminate against
noise and to insure detection.  The most sensitive instruments in the
worldwide network are associated with the Laser Interferometer
Gravitational-wave Observatory, or LIGO.  LIGO has a pair of four
kilometer, L-shaped interferometers located in Hanford, Washington and
Livingston, Louisiana.  Closely associated with LIGO is GEO600, a 600
meter interferometer near Hannover, Germany.  Because of its shorter
arms, GEO cannot achieve the same sensitivity as the LIGO detectors.
However, it has been used as a testbed for advanced interferometry
techniques, which has allowed it to maintain its role as an important
part of the worldwide detector network.  Completing the present
network is Virgo, a three kilometer interferometer located in Pisa,
Italy, and operated by a French-Italian collaboration.  Its
sensitivity is fairly close to that of the LIGO instruments.
Discussion of recent performance and upgrade plans for these three
instruments can be found here {\cite{ligogeovirgo}}.

A source of mass $M$ and size $R$ has a natural GW frequency of $f
\sim (1/2\pi)\sqrt{GM/R^3}$.  A compact source has size $R \sim
\mbox{several} \times GM/c^2$.  For such sources, the natural GW
frequency is in the high-frequency band if $M \sim 1 - 100\,M_\odot$.
For this reason, the high-frequency band largely targets objects like
neutron stars and black holes.  One of the most important sources in
this band is the coalescence of binary neutron star systems ---
essentially, the last several minutes of systems like the Hulse-Taylor
binary pulsar.  Binaries containing black holes may also be important
sources, though our poorer understanding of the formation of compact
binaries with black holes make their rates substantially less certain.

As mentioned above, the LIGO and Virgo instruments are presently
undergoing an upgrade to ``advanced'' sensitivity, which will give
them a reach to binary neutron star inspiral of about 200 Mpc.  This
is far enough that astrophysical models suggest they should measure
multiple coalescence events per year {\cite{NSNSrate}}.  The rate for
events involving black holes could plausibly be even higher: the
signal from black hole binaries is stronger, greatly increasing the
observable distance (and hence sensitive volume) {\cite{fh1998}}.  The
LIGO instruments are expected to begin observations at the first stage
of advanced sensitivity in 2016 (see discussion of detector
commissioning timetables and associated references in
Ref.\ {\cite{adv_obs}}), and should reach their final advanced design
by 2018.  Virgo is expected to follow LIGO by about two years.

\subsection{Low frequency}
\label{sec:low}

The low frequency band extends from as low as $10^{-5}$ Hz up to about
$1$ Hz, and is targeted using laser interferometry between spacecraft.
This band is particularly source rich.  Low frequency GW detectors are
expected to measure signals from dozens of coalescing massive binary
black holes {\cite{lisambbh}} (similar to the binaries targeted by
pulsar timing arrays, though at lower masses and at the tail end of
the GW-driven inspiral); from dozens to possibly hundreds of stellar
mass compact objects captured onto strong-field orbits of $\sim
10^6\,M_\odot$ black holes {\cite{emri}}; and from millions of close
binary star and binary white-dwarf systems in our galaxy
{\cite{cwdb}}.  There may even be strong signals from processes
related to phase transitions in the early universe: if the electroweak
transition occuring at temperatures of a few TeV is first order (as
some scenarios for baryogenesis suggest it could be {\cite{mrm}}),
then we expect a stochastic background signal peaked at $f \sim
1\,{\rm mHz}(T/{\rm TeV})$ from collisions of domain walls associated
with the transition.

The promise of this band has been known for quite some time, and has
motivated several proposed missions to measure GWs at these
frequencies.  From the late 1990s until early 2011, the focus was
LISA, the Laser Interferometer Space Antenna.  LISA was proposed as a
joint ESA-NASA mission, consisting of a three-spacecraft constellation
orbiting the sun in an equilateral triangle with sides of $5\times
10^6$ km.  Each spacecraft was to be placed into an orbit such that
the constellation orbited the sun once per year, lagging the Earth by
$20^\circ$, and inclined $60^\circ$ with respect to the ecliptic.  By
measuring the separation between drag-free proof masses in the
spacecraft using phase-locked laser transponders with picometer
accuracy, LISA would have achieved sufficient sensitivity to measure a
rich spectrum of sources in this band over a multiyear mission
lifetime.  See Ref.\ {\cite{lowfwp}} and references therein for
detailed discussion.

Sadly for those of us in the United States, funding constraints have
forced NASA to withdraw from this mission.  The European LISA partners
have forged ahead with plans for eLISA (``evolved LISA'',
{\cite{elisawp}}).  The European Space Agency has selected ``The
Gravitational Universe'' as the science theme for their L3 launch
opportunity, which is currently scheduled for 2034; eLISA is the
leading mission concept under development to implement this theme.
The design of eLISA can be expected to evolve in the next decade or
so, but the present design envisions a somewhat smaller LISA-like
constellation ($10^6$ km arms), with most likely a shorter mission
lifetime.  This design should achieve an impressive fraction of the
original LISA source science {\cite{elisawp}}.  Within the US, NASA's
Physics of the Cosmos Program Advisory Group (PhysPAG
{\cite{physpag}}) formed a study group {\cite{gwsig}} to evaluate what
options might be possible should budgets allow NASA to rejoin a
space-based GW mission (perhaps after the launch of the James Webb
Space Telescope).  Options being considered range from junior partner
with ESA in eLISA to the development of NASA-only mission similar to
eLISA (for example, ``SGO-Mid'' {\cite{sgomid}}, the middle range of a
suite of Space-based Gravitational-wave Observatories that were
examined in a study of possible GW missions).

\subsection{Very low frequency}
\label{sec:very}

Very low frequency GWs are targeted by timing of pulsars.  This
technique uses the fact that millisecond pulsars are very precise
clocks; indeed, the stability of some pulsars rivals laboratory atomic
clocks.  Using a well-characterized millisecond pulsar as the light
source and a radio telescope on the Earth as the sensor, this
technique implements Bondi's idea for measuring gravitational waves in
the band from roughly $10^{-9}-10^{-7}\,{\rm Hz}$.  These boundaries
of this band are set by practical considerations: One must integrate a
pulsar's signal for a few months (i.e., a time $\sim 10^7$ seconds) in
order for a GW signal to stand above the expected noise level; and
data on pulsars that are best suited to this analysis only goes back a
few decades ($\sim 10^9$ seconds).

In this frequency band, the two most plausible sources are the
coalescence of massive binary black holes, and a high-frequency tail
of the primordial GWs described in Sec.\ {\ref{sec:ultra}}.  We will
defer discussion of this tail of primordial GWs to
Sec.\ {\ref{sec:echoes}}, and briefly describe here GWs from binaries
containing massive black holes.  Such binaries are formed by the
merger of galaxies which themselves have massive black holes at their
cores.  Population synthesis estimates based on models of structure
formation and galaxy growth suggest there should be a substantial
population of such binaries whose members are black holes of $10^6 -
10^8\,M_\odot$.  The GWs produced by these binaries combine to form a
stochastic background in the very low frequency band {\cite{ptambbh}}.
This background is targeted by pulsar timing observations.

In the past several years, the promise of measuring this background
has motivated the formation of three collaborations to precisely time
a large number of pulsars to measure this background: NANOGrav, the
North American Nanohertz Obsevatory for Gravitational Waves
{\cite{nanograv}}; EPTA, the European Pulsar Timing Array
{\cite{epta}}; and PPTA, the Parkes Pulsar Timing Array {\cite{ppta}}.
These three collaborations together form IPTA, the International
Pulsar Timing Array.  They are presently timing about 40 pulsars, and
have set upper limits on a background of GWs in the nanohertz
frequency band {\cite{ptalimits}}.  As they find additional pulsars
that are ``good timers'' and build a longer baseline of timing data,
these limits will grow stronger, and either begin cutting into
predictions from galaxy formation and growth models (which will begin
to limit the space of possible binary formation models
{\cite{ptalimits}}), or produce a detection in this band.

\subsection{Ultra low frequency}
\label{sec:ultra}

The ultra low frequency GW band, $10^{-13}\,{\rm Hz} \lesssim f
\lesssim 10^{-18}\,{\rm Hz}$, is best described using wavelength: it
consists of GWs with $c/H_0 \gtrsim \lambda \gtrsim 10^{-5} c/H_0$.
In other words, these are waves that vary on lengthscales comparable
to the size of our universe.  To make strong GWs with quadrupole
dynamics on these scales would require relativistic masses that
stretch across much of the sky.  Such masses would upset the observed
homogeneity of the universe on these scales.  Some mechanism other
than quadrupole dynamics must be invoked to describe these GWs.

Such a mechanism is provided by cosmic inflation {\cite{inflation}},
the hypothesized epoch of false-vacuum-driven expansion when the
universe repeatedly doubled in size, with a doubling time of $\sim
10^{-37}$ seconds.  For an intuitive picture of how inflation does
this, consider the following argument due to Allen {\cite{bruce}}.
Consider a ground state quantum simple harmonic oscillator in 1-D,
with potential $V = mx^2\omega_i^2/2$.  This system's wavefunction is
\begin{equation}
\psi \equiv \psi^i_0(x) =
\left(\frac{m\omega_i}{\pi\hbar}\right)^{1/4}\exp\left(-\frac{m\omega_i
  x^2}{2\hbar}\right)\;.
\end{equation}
Now imagine that the potential very suddenly changes to $V =
mx^2\omega_f^2/2$.  The change is so rapid that the wavefunction cannot
adiabatically evolve with the potential; indeed, to first
approximation, the wavefunction is left unchanged.  However, it is no
longer a ground state, but is instead a highly excited state of the
final potential.  To see this, we write
\begin{equation}
\psi = \sum_{n = 0}^\infty c_n \psi^f_n(x)\;,
\end{equation}
where $\psi^f_n = N^f_n H_n(x\sqrt{m\omega_f/\hbar})\exp(-m\omega_f
x^2/2\hbar)$ are basis functions corresponding to states of the final
potential; $H_n$ is a Hermite polynomial.  A straighforward exercise
yields the expansion coefficients $c_n$, from which we deduce the
energy of the final state to be
\begin{equation}
E = \hbar\omega_f\left(\frac{1}{2} + \frac{(\omega_f -
  \omega_i)^2}{4\omega_i\omega_f}\right) \simeq
\hbar\omega_f\left(\frac{1}{4}\frac{\omega_i}{\omega_f}\right)\;.
\end{equation}
(We use $\omega_i \gg \omega_f$ in the last step.)  The change in the
potential created $N = \omega_i/4\omega_f$ quanta.

With this cartoon-level sketch in mind, consider now cosmic inflation.
Prior to inflation, the universe is in a vacuum state, filled with the
ground state of various fields, including gravity.  Inflation acts
like the suddenly changing potential, sharply reducing the frequencies
associated with modes of the field, doing so rapidly enough that the
evolution is non-adiabatic.  GWs in particular are created by this
process; this is the only known mechanism for producing GWs with
wavelengths near the Hubble scale while maintaining the homogeneity
and isotropy of the universe.  For more detailed discussion that goes
beyond this heuristic picture, see Refs.\ {\cite{bruce,kolbturner}}.

Following inflation, the GWs that are produced by this process
propagate through the universe.  Because of gravity's weakness, they
barely interact with matter as they propagate, just stretching and
squeezing the primordial plasma in the young expanding universe.  In
particular, the GWs stretch and squeeze the plasma at the moment of
recombination, when the plasma has cooled enough that atoms can form,
and photons begin to free stream, forming the cosmic microwave
background (CMB).  This stretching and squeezing creates a quadrupolar
temperature anisotropy in the plasma at recombination, which causes
the CMB to be linearly polarized {\cite{cmbpol}}.  The GWs thus leave
an imprint on the cosmic microwave background.

Other processes polarize the CMB as well.  In particular, the density
inhomogeneities primarily responsible for the famous temperature
fluctuations in the CMB also cause quadrupolar anisotropies that lead
to linear polarization.  One can however detangle these two sources of
polarization in a model-independent fashion.  Polarization is a
vector, and can be written as the gradient of a scalar potential plus
the curl of a vector potential.  The contributions from the gradient
of the scalar potential are known as ``E modes,'' and those from the
curl of the vector potential as ``B modes.''  Because density
perturbations have no handedness associated with them, they can only
create E modes.  GWs can have a handedness, and so they can source
both E and B modes.  The B modes are thus a unique and powerful
signature of primordial GWs.  Since inflation is the only mechanism we
know of to create GWs with wavelengths close to the Hubble length,
their detection is considered to be a ``smoking gun'' for cosmic
inflation.

Prior to 17 March 2014, the standard lore was that these GWs were in
all likelihood so weak that we were quite some time away from
measurement of these waves.  Bounds inferred from the temperature
spectrum by the WMAP and Planck satellites {\cite{WMAP,Planck}} were
pointing to relatively small levels of primordial GWs; also,
foreground effects, which can transform an E-mode signal into a B-mode
{\cite{emodelensing}}, were throught to be potentially quite daunting.
It was thus quite stunning\footnote{When I originally presented this
  material at the September 2013 TAUP Conference, I gave the standard
  line that these GWs would like require years of study to understand
  foregrounds and other systematic effects before any discovery.
  Fortunately, my tendency to procrastinate meant that I didn't write
  up this article until well after the BICEP2 announcement.  This has
  given me a chance to wipe a little bit of egg off my face.}  when
the BICEP2 collaboration announced a 7-$\sigma$ detection of B-modes
from their telescope at the South Pole {\cite{bicep2}}.  This result
needs to be confirmed by other experiments, and it must be understood
why they are (apparently) in discord with previous upper limits.
Their results suggest a GW strength high enough that confirmation
should be likely fairly quickly.

As we write this article, the BICEP2 announcement is only a month old.
We can expect a lot of work in this field, with (hopefully)
confirmation very soon, and future work allowing us to begin probing
the nature of inflation directly.  We conclude this section by noting
that, if confirmed, the BICEP2 result represents the first time that
the influence of GWs on matter (other than the waves' own source) has
been measured.  This will be the first of many examples of GWs being
exploited for astronomy and cosmology.

\section{Cosmology with gravitational-wave measurements}
\label{sec:cosmo}

When the author was asked to speak at the 2013 TAUP meeting, the
invitation requested a review of GWs and cosmology.  Any review of
this subject, prior to the direct detection of GWs, necessarily must
be speculative to some degree.  The reader should take the discussion
here to indicate what cosmological applications of GW physics have
been seriously thought about to date.  It will be interesting (and
possibly amusing) to compare this discussion with the applications
that actually develop once GW measurement becomes routine.

\subsection{Standard sirens}
\label{sec:sirens}

Chief among the sources across several GW bands are the inspiral and
merger of compact binary sources.  A binary is a nearly perfect
quadrupole radiator and, unless general relativity fails in the deep
strong field, its waves have a form that depends only on physical
parameters of the system.  The waveform depends most strongly on the
source binary's masses and spins, the angles which determine its
position on the sky and orientation with respect to the line of sight,
and the distance to the source.  Schematically, a measured binary
waveform takes the form
\begin{equation}
h_{\rm meas} = \frac{G(1+z){\cal M}/c^2}{D_L(z)}[\pi(1 + z){\cal
    M}f(t)]^{2/3}{\cal
  F}\left(\mbox{angles}\right)\cos\Phi\left(m_1,m_2,\vec S_1, \vec
S_2;t\right)\;.
\label{eq:waveform_schematic}
\end{equation}
The binary's masses and spins strongly affect the waveform's phase
evolution $\Phi$ (note that $f = (1/2\pi)d\Phi/dt$).  Because data
analysis is based on phase coherently matching data to a model, $\Phi$
typically will be measured to within a fraction of a radian.  The
masses and spins can thus be determined to good accuracy (where
details of ``good'' depend on the measurement's signal-to-noise ratio,
and how well certain near degeneracies between parameters are broken;
see {\cite{massspin}} for more detailed discussion).

The mass parameter ${\cal M} \equiv (m_1m_2)^{3/5}/(m_1+m_2)^{1/5}$,
known as the ``chirp mass,'' is determined extremely well.  The wave's
amplitude effectively depends only on the angles which determine the
binary's orientation and sky position, and on the binary's luminosity
distance $D_L(z)$.  If there are enough GW detectors to measure both
GW polarizations, then inclination angles can be determined.  If it is
likewise possible to determine sky position, then {\it all} of the
important source angles are determined.  {\it In this case, measuring
  the binary's waveform directly determines its luminosity distance
  $D_L$.}  Binary inspiral thus has the potential to act as a standard
siren\footnote{In most astronomical applications, this would be called
  a standard candle.  However, in many respects, GWs can be regarded
  as sound-like, and the use of ``siren'' rather than ``candle'' has
  been adopted to reflect this.} --- a precisely calibrated source
whose measured characteristics encode distance to the source
{\cite{schutz,holzhughes}}.

In the high-frequency band, standard sirens are likely to be
coalescing binaries, measured out to a distance of a few hundred Mpc.
These events may be accompanied by an ``electromagnetic'' counterpart
{\cite{nsns_em}}, which opens the possibility that they could be
measured simultaneously in GWs and with telescopes in various
wavebands {\cite{adv_obs}}.  In the low-frequency band, standard
sirens are likely to be merging black holes out to a redshift $z \sim
1$ --- perhaps even further.  All of these measurements promise a new
way of pinning down cosmic expansion {\cite{schutz,samaya2}} in a
manner which would have total different systematics from other
techniques.

\subsection{Tracing massive black hole growth}
\label{sec:tracingbhs}

It appears that the galaxies which populate our universe grew in a
hierarchical manner, through the repeated merger of smaller galaxies
or protogalaxies.  At some point in this process, black holes formed
in the cores of at least some of these structures.  When these
galaxies or protogalaxies merge, the black holes will eventually come
close enough to one another to bind into a tight binary that evolves
through GW emission (although multiple evolutionary steps are needed
to reach the point that GW emission is important {\cite{bbr}}).  The
GW signal from such black holes (with masses from $10^5 -
10^7\,M_\odot$) can be measured by space interferometers like LISA and
eLISA to high redshift.  In many cases, enough signal will be measured
(many months or even a few years of inspiral, through to final merger
into a single black hole) that the system's parameters can determine
the rest frame masses and spins, as well as the source redshift
{\cite{untangling}}.

Black holes are completely determined by their masses and spins
{\cite{nohair}}.  If we determine these two parameters, we have
determined everything that can be known about them.  How a black
hole's mass and spin evolves is quite sensitive to the details of how
the black hole gains mass: accretion tends to spin up black holes, and
mergers tend to spin them down (with significant variation depending
on the detailed mode in which accretion is presumed to operate).
Precision data on merging black holes' masses and spins over a range
of redshifts will provide a tremendous amount of information
clarifying how black holes formed and grew from very early cosmic
epochs {\cite{lisambbh}}.

As discussed in Sec.\ {\ref{sec:low}}, observations of these merging
black holes in the low frequency band are quite some time in the
future.  Fortunately, very low frequency observations with pulsar
timing arrays are likely to begin telling us about a related
population of merging black holes relatively soon: the prime source
for these arrays are merging massive black holes which likewise form
from the merger of galaxies {\cite{ptambbh}}.  These black hole
binaries differ in several important ways from those targeted by space
interferometry: they are at rather higher masses than the targets of
interferometers ($\sim 10^8\,M_\odot$ rather than $\sim
10^6\,M_\odot$); they typically come from much lower redshift; and
they involve binaries that are millions of years away from their final
merger.  However, they are similar in that the measurement of these
waves directly probes a dynamical consequence of galaxy assembly and
evolution.  Recent papers make it clear that there is much that can be
learned by a discovery of GWs in this band (e.g.,
{\cite{sean,alberto}}).

\subsection{Echoes from the early universe}
\label{sec:echoes}

Finally, there is much that can be learned from GWs produced in the
early universe.  We have already described the process by which
inflation produces GWs, and are eagerly waiting for confirmation of
the BICEP2 results announced in Ref.\ {\cite{bicep2}}.  If confirmed,
it will soon be possible to measure this spectrum at different scales,
making it possible to begin probing the detailed physics of the
inflationary potential.  It will then be possible to begin
phenomenology of processes at the roughly $10^{15}$ GeV scale
associated with inflation.

It is worth noting here that inflation does not just produce GWs near
the Hubble scale, but yields a very broad-band spectrum of
fluctuations.  A very simple estimate predicts a flat spectrum from
about\footnote{This frequency is related to the transition from the
  post-inflationary radiation-dominated universe to a matter-dominated
  universe {\cite{bruce,kolbturner}}.} $10^{-15}$ Hz to well-above the
high frequency band.  A more careful analysis shows that the spectrum
actually rolls off at high frequencies, with a value that depends on
$n_T$, the spectral index of tensor modes {\cite{turner96}}.  For the
purposes of this article, this high-frequency tail produces waves that
are well below the projected sensitivity of any measurement that is
foreseeable in the next decade or two.  If the amplitude of GWs found
by BICEP2 is confirmed, then there will be a very strong case to begin
developing experiments or missions to measure this background (for
example, BBO {\cite{bbo}} or DECIGO {\cite{decigo}}).

Finally, we reiterate that although inflation is guaranteed to produce
GWs, there are other early universe processes that could produce such
a signal.  Perhaps the most interesting possibility is that of GWs
from a first-order electroweak phase transition {\cite{mrm}},
discussed in Sec.\ {\ref{sec:very}}.  Such a signal would require
space interferometry, but success in other parts of the GW spectrum
will strengthen the case for such a mission.

\section{Outlook}
\label{sec:outlook}

For the past two or so decades, GW has been described as a field of
great promise.  It is now on the threshold of delivering on that
promise.  If the BICEP2 results are soon confirmed, the first delivery
has in fact already arrived.  With advanced ground-based detectors
soon to begin operations, and with pulsar timing arrays continuing to
advance in capability, we can expect to begin using information from
three of the four major GW frequency bands in the next several years.
The fourth band will probably take somewhat longer (anything involving
space missions involves a long lead time), but solid detections in the
other bands will build enthusiasm for probing the rich low-frequency
band's datastream.

The decade of gravitational-wave discovery has begun.

\section*{Acknowledgments}

I thank the organizers of TAUP2013, particularly Wick Haxton, for
inviting me to speak, and for repeatedly extending the deadline for
this proceedings article.  I also thank Scott Ransom and Ingrid Stairs
for helping me track down good references summarizing our knowledge of
pulsars in inspiraling binaries, and Guido Mueller for helpful
feedback on a draft of this article.  My work on gravitational-wave
physics is supported by NSF Grant PHY-1068720.




\begin{thebibliography}{00}

\bibitem{ht} J.\ M.\ Weisberg, D.\ J.\ Nice, and J.\ H.\ Taylor,
  Astrophys.\ J.\ {\bf 722}, 1030 (2010).

\bibitem{inspiraling} J.\ Antoniadis et al, Science {\bf 340}, 448
  (2013) (see especially Table 2, which lists all known inspiraling
  pulsar-white dwarf and four known pulsar-neutron star binaries);
  R.\ D.\ Ferdman et al, Astrophys.\ J.\ {\bf 767}, 85 (2013)
  (especially Table 1, which lists all known double neutron star
  systems with precisely measured components masses).

\bibitem{wd_inspiral} J.\ J.\ Hermes et al, Astrophys.\ J.\ {\bf 757},
  21 (2012); see also W.\ R.\ Brown, M.\ Kilic, C.\ Allende Prieto,
  A.\ Gianninas, and S.\ J.\ Kenyon, Astrophys.\ J.\ {\bf 769}, 66
  (2013).

\bibitem{bondi} H.\ Bondi, Nature {\bf 179}, 1072 (1957).

\bibitem{hughesthorne} S.\ A.\ Hughes and K.\ S.\ Thorne,
  Phys.\ Rev.\ D {\bf 58}, 122002 (1998).

\bibitem{ligogeovirgo} S.\ J.\ Waldman for the LIGO Science
  Collaboration, proceedings of the 2010 Recontres de Blois conference
  and arXiv:1103.2728; H.\ L\"uck et al, J.\ of Phys.:
  Conf.\ Ser.\ {\bf 228}, 012012 (2012); M.-A.\ Bizouard and
  M.\ A.\ Papa, Comptes Rendus Physique {\bf 14}, 352 (2013); the LIGO
  Scientific Collaboration and the Virgo Collaboration,
  arXiv:1203.2674.

\bibitem{NSNSrate} J.\ Abadie et al, Class.\ Quantum Grav.\ {\bf 27},
  173001 (2010).

\bibitem{fh1998} E.\ E.\ Flanagan and S.\ A.\ Hughes, Phys.\ Rev.\ D
  {\bf 57}, 4535 (1998).

\bibitem{adv_obs} L.\ P.\ Singer et al, Astrophys.\ J., submitted;
  arXiv:1404.5623.

\bibitem{lisambbh} A.\ Sesana, M.\ Volonteri, and F.\ Haardt,
  Mon.\ Not.\ R.\ Astron.\ Soc.\ {\bf 377}, 1711 (2007); T.\ Tanaka
  and Z.\ Haiman, Astrophys.\ J.\ {\bf 696}, 1798 (2009); A.\ Sesana,
  J.\ Gair, E.\ Berti, and M.\ Volonteri, Phys.\ Rev.\ D {\bf 83},
  044036 (2011).

\bibitem{emri} J.\ R.\ Gair, Class.\ Quantum Grav.\ {\bf 26}, 094034
  (2009).

\bibitem{cwdb} T.\ R.\ Marsh, Class.\ Quantum Grav.\ {\bf 28}, 094019
  (2011).

\bibitem{mrm} D.\ E.\ Morrissey and M.\ J.\ Ramsey-Musolf,
  New.\ J.\ Phys.\ {\bf 14}, 125003 (2012).

\bibitem{lowfwp} T.\ Prince for the LISA International Science Team,
  white paper submitted to the 2010 Decadal Survey in Astronomy and
  Astrophysics, available at {\tt
    http://sites.nationalacademies.org/bpa/BPA\_050603}

\bibitem{elisawp} P.\ Amaro Seoane et al, {\it The Gravitational
  Universe}, a white paper describing the science theme addressed by
  the proposed eLISA gravitational-wave mission; arXiv:1305.5720.

\bibitem{physpag} {\tt http://pcos.gsfc.nasa.gov/physpag/}

\bibitem{gwsig} See {\tt http://pcos.gsfc.nasa.gov/sigs/gwsig.php} for
  further discussion.

\bibitem{sgomid} R.\ Stebbins, American Astronomical Society, 219th
  AAS Meeting, 146.24 (2012).  Further information and discussion at
  {\tt
    http://pcos.gsfc.nasa.gov/studies/gravitational-wave-mission.php}

\bibitem{bruce} B.\ Allen, in {\it Relativistic gravity and
  gravitational radiation}, eds.\ J.-A.\ Marck and J.-P.\ Lasota,
  Proceedings of the Les Houches School on AStrophysical Sources of
  Gravitational Waves, Cambridge University Press, Cambridge, 1996;
  also gr-qc/9604033.

\bibitem{kolbturner} E.\ W.\ Kolb and M.\ S.\ Turner, {\it The Early
  Universe} (Addison-Wesley, 1990).

\bibitem{ptambbh} P.\ A.\ Rosado and A.\ Sesana,
  Mon.\ Not.\ R.\ Astron.\ Soc.\ {\bf 439}, 3986 (2014).

\bibitem{nanograv} {\tt http://nanograv.org}

\bibitem{epta} {\tt http://www.upta.eu.org}

\bibitem{ppta} {\tt http://www.atnf.csro.au/research/pulsar/array/}

\bibitem{ptalimits} R.\ van Haasteren et al,
  Mon.\ Not.\ R.\ Astron.\ Soc.\ {\bf 414}, 3117 (2011);
  R.\ M.\ Shannon et al, Science {\bf 342}, 334 (2013);
  Z.\ Arzoumanian et al, Astrophys.\ J., submitted, and
  arXiv:1404.1267.

\bibitem{inflation} A.\ Guth, Phys.\ Rev.\ D {\bf 23}, 347 (1981);
  A.\ D.\ Linde, Phys.\ Lett.\ {\bf 108B}, 389 (1982); A.\ Albrecht
  and P.\ J.\ Steinhardt, Phys.\ Rev.\ Lett.\ {\bf 48}, 1220 (1982).

\bibitem{cmbpol} J.\ R.\ Bond and G.\ Efstathiou,
  Astrophys.\ J.\ Lett.\ {\bf 285}, L47 (1984); A.\ G.\ Polnarev,
  Soc.\ Astron.\ {\bf 29}, 607 (1985); U.\ Seljak and M.\ Zaldarriaga,
  Phys.\ Rev.\ Lett.\ {\bf 78}, (1997); M.\ Kamionkowski, A.\ Kosowsky,
  and A.\ Stebbins, Phys.\ Rev.\ Lett.\ {\bf 78}, 2058 (1997).

\bibitem{WMAP} G.\ Hinshaw et al, Astrophys.\ J.\ Suppl.\ {\bf 208},
  19 (2013).

\bibitem{Planck} Planck collaboration: P.\ A.\ R.\ Ade et al,
  arXiv:1303.5082.

\bibitem{emodelensing} M.\ Zaldarriaga and U.\ Seljak, Phys.\ Rev.\ D
  {\bf 58}, 023003 (1998).

\bibitem{bicep2} BICEP2 Collaboration, P.\ A.\ R.\ Ade et al,
  arXiv:1403.3985.

\bibitem{massspin} R.\ N.\ Lang and S.\ A.\ Hughes, Phys.\ Rev.\ D
  {\bf 74}, 122001 (2006); R.\ N.\ Lang, S.\ A.\ Hughes, and
  N.\ J.\ Cornish, Phys.\ Rev.\ D {\bf 84}, 022002 (2011).

\bibitem{schutz} B.\ F.\ Schutz, Nature {\bf 323}, 310 (1986).

\bibitem{holzhughes} D.\ E.\ Holz and S.\ A.\ Hughes,
  Astrophys.\ J.\ {\bf 629}, 15 (2005).

\bibitem{nsns_em} B.\ D.\ Metzger and E.\ Berger, Astrophys.\ J.\ {\bf
  746}, 48 (2012).

\bibitem{samaya2} S.\ Nissanke, D.\ E.\ Holz, N.\ Dalal,
  S.\ A.\ Hughes, J.\ L.\ Sievers, and C.\ M.\ Hirata, Astrophys.\ J.,
  submitted; arXiv:1307.2638.

\bibitem{bbr} M.\ C.\ Begelman, R.\ D.\ Blandford, and M.\ J.\ Rees,
  Nature {\bf 287}, 307 (1980).

\bibitem{untangling} S.\ A.\ Hughes,
  Mon.\ Not.\ R.\ Astron.\ Soc.\ {\bf 331}, 805 (2002).

\bibitem{nohair} W.\ Israel, Phys.\ Rev.\ {\bf 164}, 1776 (1967);
  B.\ Carter, Phys.\ Rev.\ Lett.\ {\bf 26}, 331 (1971);
  D.\ C.\ Robinson, Phys.\ Rev.\ Lett.\ {\bf 34}, 905 (1975).

\bibitem{sean} S.\ T.\ McWilliams, J.\ P.\ Ostriker, and
  F.\ Pretorius, Astrophys.\ J., submitted; arXiv:1211.5377.

\bibitem{alberto} A.\ Sesana, Mon.\ Not.\ R.\ Astron.\ Soc.\ {\bf
  433}, L1 (2013).

\bibitem{turner96} M.\ S.\ Turner, Phys.\ Rev.\ D {\bf 55}, 435
  (1997).

\bibitem{bbo} E.\ S.\ Phinney et al, {\it Big Bang Observer Mission
  Concept Study} (NASA); see also C.\ Cutler and D.\ E.\ Holz,
  Phys.\ Rev.\ D {\bf 80}, 104009 (2009).

\bibitem{decigo} S.\ Kawamura et al, Class.\ Quantum Grav.\ {\bf 28},
  094011 (2011).

\end{thebibliography}
\end{document}